\pdfoutput=1

\documentclass[aps,prl,reprint,showkeys,superscriptaddress]{revtex4-1}
\usepackage{amsmath}
\usepackage{bm}
\usepackage{appendix}
\usepackage{graphicx}
\usepackage[pdftex,colorlinks,hyperindex,plainpages=false,bookmarksopen,bookmarksnumbered,pdfusetitle,allcolors={blue}]{hyperref}
\usepackage{float}
\usepackage{mathrsfs}
\usepackage{bm}
\usepackage{bbold}
\usepackage{cases}
\usepackage{comment}

\newcommand\dd{\text{d}}

\begin{document}
\title{Electrical and thermal transport in antiferromagnet--superconductor junctions}

\author{Martin F.~Jakobsen}
\affiliation{Center for Quantum Spintronics, Department of Physics, Norwegian University of Science and Technology, NO-7491 Trondheim, Norway}
\author{Kristian B.~Naess}
\affiliation{Center for Quantum Spintronics, Department of Physics, Norwegian University of Science and Technology, NO-7491 Trondheim, Norway}
\author{Paramita Dutta}
\affiliation{Department of Physics and Astronomy, Uppsala University, Box 516, S-751 20 Uppsala, Sweden}
\altaffiliation[Present address]{}
\affiliation{Institute of Physics, Sachivalaya Marg, Bhubaneswar 751005, India}

\author{Arne Brataas}
\affiliation{Center for Quantum Spintronics, Department of Physics, Norwegian University of Science and Technology, NO-7491 Trondheim, Norway}
\author{Alireza Qaiumzadeh}
\affiliation{Center for Quantum Spintronics, Department of Physics, Norwegian University of Science and Technology, NO-7491 Trondheim, Norway}

\begin{abstract}
We demonstrate that antiferromagnet--superconductor (AF--S) junctions show qualitatively different transport properties than normal metal--superconductor (N--S) and ferromagnet--superconductor (F--S) junctions. We attribute these transport features to the presence of two new scattering processes in AF--S junctions, i.e., specular reflection of holes and retroreflection of electrons. Using the Blonder--Tinkham--Klapwijk formalism, we find that the electrical and thermal conductance depend nontrivially on antiferromagnetic exchange strength, voltage, and temperature bias. Furthermore, we show that the interplay between the N\'eel vector direction and the interfacial Rashba spin-orbit coupling leads to a large anisotropic magnetoresistance. The unusual transport properties make AF--S interfaces unique among the traditional condensed-matter-system-based superconducting junctions.
\end{abstract}
\maketitle

\textit{Introduction}.-- Heterostructures composed of superconductors and nonsuperconducting materials exhibit technologically relevant quantum phenomena~\cite{Yan2018,PhysRevLett.86.2427,Robinson59,PhysRevLett.90.137003,PhysRevLett.105.077001,Baek2014,Eschrig_2015,RevModPhys.76.323}. Examples include superconducting qubits~\cite{DiCarlo2009,Mooij1036,Arute2019}, microwave resonators~\cite{129302}, single-photon detectors~\cite{doi:10.1063/1.1388868}, and AC Josephson junction lasers~\cite{Cassidy939}. Superconducting heterostructures also form the basis for experimental methods such as point--contact spectroscopy~\cite{PhysRevLett.81.3247,Soulen85,PhysRevLett.83.1427} and scanning tunneling spectroscopy~\cite{MESERVEY1994173,Bode_2003}, enabling the determination of the superconducting gap and investigations of the phase diagram in unconventional superconductors~\cite{GONNELLI201372,PhysRevB.83.094507,Daghero_2011}.

The simplest superconducting heterostructure is a normal metal (N)--superconductor (S) junction. The low bias transport is dominated by Andreev reflection (AR)~\cite{RevModPhys.76.323,Daghero_2010,Lee_2016}. In conventional AR, an incident electron is retroreflected as a hole of the opposite spin, and a Cooper pair is transmitted into the S layer. Since the Cooper pair carries a charge of $2e$ and zero heat, AR enhances electrical conductance and suppresses thermal conductance~\cite{PhysRevB.25.4515,PhysRevB.48.15198,Lee2019}. In a Josephson junction (S--N--S)~\cite{JOSEPHSON1962251}, AR can occur repeatedly, resulting in Andreev bound states that carry a supercurrent across the junction. Josephson junctions enable technologies such as electrical and thermal interferometers~\cite{Giazotto2012,PhysRevB.55.3849}.

The spin dependence of AR at superconducting interfaces causes the transport properties to change drastically when ferromagnetic (F) layers are introduced~\cite{Linder2015}. The exchange interaction splits the majority and minority spin bands in the F layer, which reduces the AR amplitude and consequently the conductance in a F--S junction~\cite{PhysRevLett.74.1657,PhysRevB.83.094507}. However, finite spin-orbit coupling (SOC) at the interface enables tunable anisotropic spin-flipped AR, which can increase the electric and thermal conductance~\cite{PhysRevLett.115.116601,PhysRevB.100.060507,PhysRevB.96.115404,dutta2020thermoelectricity}. S--F--S Josephson junctions have been shown to exhibit spin-triplet pairing, potentially enabling superconducting spin currents and qubits~\cite{PhysRevLett.104.137002,Robinson59,PhysRevB.82.060505,PhysRevLett.95.097001,RevModPhys.77.935,PhysRevLett.86.2427,BuzdinFeb1982}. However, the finite net magnetization of ferromagnets in superconducting spintronics presents a significant drawback for applications in nanoscale devices.

Antiferromagnets (AFs) are magnetically ordered materials with zero net magnetization and negligible stray fields, as well as intrinsic high-frequency dynamics. Thus, AFs are promising candidates for novel high-density and ultrafast spintronic-based nanodevices \cite{Review-AFM1}.
Based on these characteristics and recent experimental developments, the emerging field of antiferromagnetic spintronics has attracted intensive interest~\cite{MetallicAFMs,Jungwirth2016,Park2011,PhysRevLett.108.017201,PhysRevLett.113.097202,PhysRevLett.113.157201,Wadley587,Lebrun:Nature2018,Li:Nature2020,Vaidya:Science2020}. Additionally, the possible coexistence of antiferromagnetism with superconductivity \cite{PhysRevLett.91.056401, RevModPhys.76.909, PhysRevB.84.125140} shows the great potential of antiferromagnetic materials for application in superconducting spintronics \cite{Linder2015}.

AF--S junctions have been theoretically shown to exhibit additional scattering processes that differ from those in N(F)--S junctions~\cite{PhysRevLett.94.037005}. In Josephson junctions, these new scattering processes create low-energy bound states~\cite{PhysRevB.72.184510} that lead to anomalous phase shifts~\cite{PhysRevResearch.1.033095} and atomic-scale $0-\pi$ transitions~\cite{PhysRevLett.96.117005,Zhou_2019,PhysRevB.95.104513,PhysRevB.88.214512}. However, although the existence of Josephson supercurrents in S--AF--S junctions has been experimentally reported~\cite{PhysRevB.68.144517,PhysRevLett.99.017004,H_bener_2002,Constantinian2013,doi:10.1063/1.4824891,PhysRev.142.118}, other theoretical predictions have yet to be explored.

To our knowledge, the effect of these additional scattering processes on the electrical and thermal transport in AF--S bilayers remains an open question. In this paper, we address this issue and point out unique experimental signatures in the electrical and thermal conductance.

\textit{Model}.-- We consider a collinear, two-sublattice AF metal on a cubic lattice attached to a conventional $s$-wave superconductor. The AF and S are both semi-infinite and occupy the regions $z<0$ and $z> 0$, respectively. We assume a compensated interface at $z = 0$.

To investigate the electrical and thermal transport, we use the Blonder--Tinkham--Klapwijk (BTK) scattering formalism~\cite{PhysRevB.25.4515}, where the conductances are determined by the reflection coefficients of the scattering matrix. We obtain the reflection coefficients by solving the Bogoliubov--de Gennes (BdG) equation.

The BdG Hamiltonian of an AF--S junction in the continuum limit consists of a Hamiltonian for itinerant charge carriers $H_\mathrm{e}$, an antiferromagnetic exchange coupling $H_\mathrm{AF}$, an interfacial barrier potential $H_\mathrm{I}$, and a Hamiltonian modeling the S layer $H_\mathrm{S}$,
\begin{equation}
    H = H_\mathrm{e} + H_{\mathrm{AF}} +  H_{\mathrm{I}} +H_{\mathrm{S}}.
    \label{Eq:AFM-SC-Tight-Binding}
\end{equation}
The Pauli matrices $\mathbf{s}$, $\bm{\sigma}$, and $\bm{\tau}$  denote the spin, sublattice, and charge degrees of freedom, respectively. We also define $\tau_4^{\pm} = \mathrm{diag}(1,\pm K)$, where $K$ represents complex conjugation, and $\tau_{\pm} = \left(\tau_x \pm i \tau_y\right)/2$.

The Hamiltonian governing the motion of the itinerant charge carriers is~\cite{PhysRevLett.120.197202,PhysRevB.95.014403,PhysRevLett.113.157201}
\begin{equation}
    H_{\mathrm{e}} = \gamma(\mathbf{p})\, \tau_z \otimes \sigma_x \otimes s_0 - \mu\, \tau_z \otimes \sigma_0 \otimes s_0,
\end{equation}
where $\gamma(\mathbf{p}) =  \left(\mathbf{p}^2 - \hbar^2\bm{k}_0^2 \right)/2m$ is the kinetic energy, $\mathbf{p} = -i \hbar \bm{\nabla}$ is the momentum operator, $m$ is the effective mass of the charge carriers, $\bm{k}_0$ is the wavevector at which $\gamma(\hbar \mathbf{k}_0)=0$, and $\hbar$ is the reduced Planck constant. The chemical potential is $\mu = \mu_{\mathrm{AF}}\Theta(-z) + \mu_{\mathrm{S}}\Theta(z)$, where $\Theta(\cdot)$ is the Heaviside step function.

The {\it s--d} exchange interaction between localized antiferromagnetic moments and itinerant spins reads~\cite{PhysRevLett.120.197202,PhysRevB.95.014403,PhysRevLett.113.157201}
\begin{equation}
    H_{\mathrm{AF}} = J\, \tau_4^- \otimes  \sigma_z \otimes \left(\mathbf{n} \cdot \bm{s}\right),
\end{equation}
where $J = J_0 \Theta(-z)$ denotes the interaction strength and $\mathbf{n}=(\sin\theta \cos \phi, \sin \theta \sin \phi, \cos \theta)$ is the uniform N\'eel vector. We assume strong anisotropy set the direction of spins and suppresses quantum fluctuations.
The interfacial potential is,
\begin{equation}
\begin{split}
    H_{\mathrm{I}} &= V\, \tau_z \otimes \sigma_0 \otimes s_0 + \lambda_{\mathrm{R}}\, \tau_4^+ \otimes \sigma_x \otimes\left[ \left(\bm{s}\times \mathbf{q}\right)\cdot \hat{z} \right],
    \end{split}
\end{equation}
where $V = V_0 \delta(z)$ is the strength the spin-independent potential barrier and $\lambda_\mathrm{R} = \lambda_0 \delta(z)$ is the strength the Rashba SOC (RSOC) ~\cite{Bychkov_1984} due to the inversion symmetry breaking in the $z$ direction. These terms permit spin-conserving and spin-flipped reflection processes, respectively. AR and spin-flipped AR result in spin-singlet and spin-triplet Cooper pairs in S, respectively.

Finally, we model the S layer using a mean-field BCS Hamiltonian:
\begin{equation}
    \begin{split}
        H_{\mathrm{S}} = \Delta(T)\, \tau_{+} \otimes \left(\sigma_0 \otimes i s_y\right) + \mathrm{h.c.},
    \end{split}
\end{equation}
where $\Delta(T) = \Delta_0 \tanh\left(1.74 \sqrt{(T_c/T)-1}\right) \Theta(z)$ is an interpolation formula for the temperature-dependent gap of an $s$-wave superconductor with a critical temperature $T_c$~\cite{Senapati2011,PhysRevLett.116.237001,footnote}. $\Delta_0$ is the constant bulk value of the gap~\cite{PhysRevB.72.184510}.

To determine the reflection coefficients, we solve the BdG eigenvalue problem,
\begin{equation}
    H \psi = E \psi,
    \label{Eq:ContinuumBdGEquation}
\end{equation}
where $\psi$ is an eigenvector with eigenvalue $E>0$ (see the Supplemental Material~\cite{supplement}). The $x$ and $y$ directions are translationally invariant. Hence, the eigenvector takes the form $\psi =\chi e^{i \mathbf{q}_{\parallel}  \cdot \mathbf{r}} e^{i q_z z} $, where $\mathbf{q}_{\parallel} = (q_x, q_y,0)$ is the conserved component of the wavevector parallel to the interface and $q_z$ is the wavevector component normal to the interface. The spinor $\chi$ is expressed in the basis~\cite{supplement}:
\begin{equation}
    \chi = \left(A_{e\uparrow},A_{e\downarrow},B_{e\uparrow},B_{e\downarrow}, A_{h\uparrow},A_{h\downarrow},B_{h\uparrow},B_{h\downarrow}  \right).
\end{equation}
Here, $A$ ($B$), $\uparrow$ ($\downarrow$), and $e$ ($h$) refer to sublattice, spin, and charge degrees of freedom, respectively.
Substituting the eigenvector into Eq.~\eqref{Eq:ContinuumBdGEquation} for $z<0$, give the wavevectors $q_z = q_{e(h)}^{\pm}$ in the AF layer,
\begin{equation}
q_{ e(h)}^{\pm} = \sqrt{k_0^2 - \mathbf{q}_{\parallel}^2  \pm \frac{2m}{\hbar^2} \sqrt{\left(E +(-) \mu_{\mathrm{AF}}\right)^2 - J^2}}.
\label{Eq:Eigenvalues}
\end{equation}

\begin{figure}[h]
\centering
\includegraphics[width=\columnwidth]{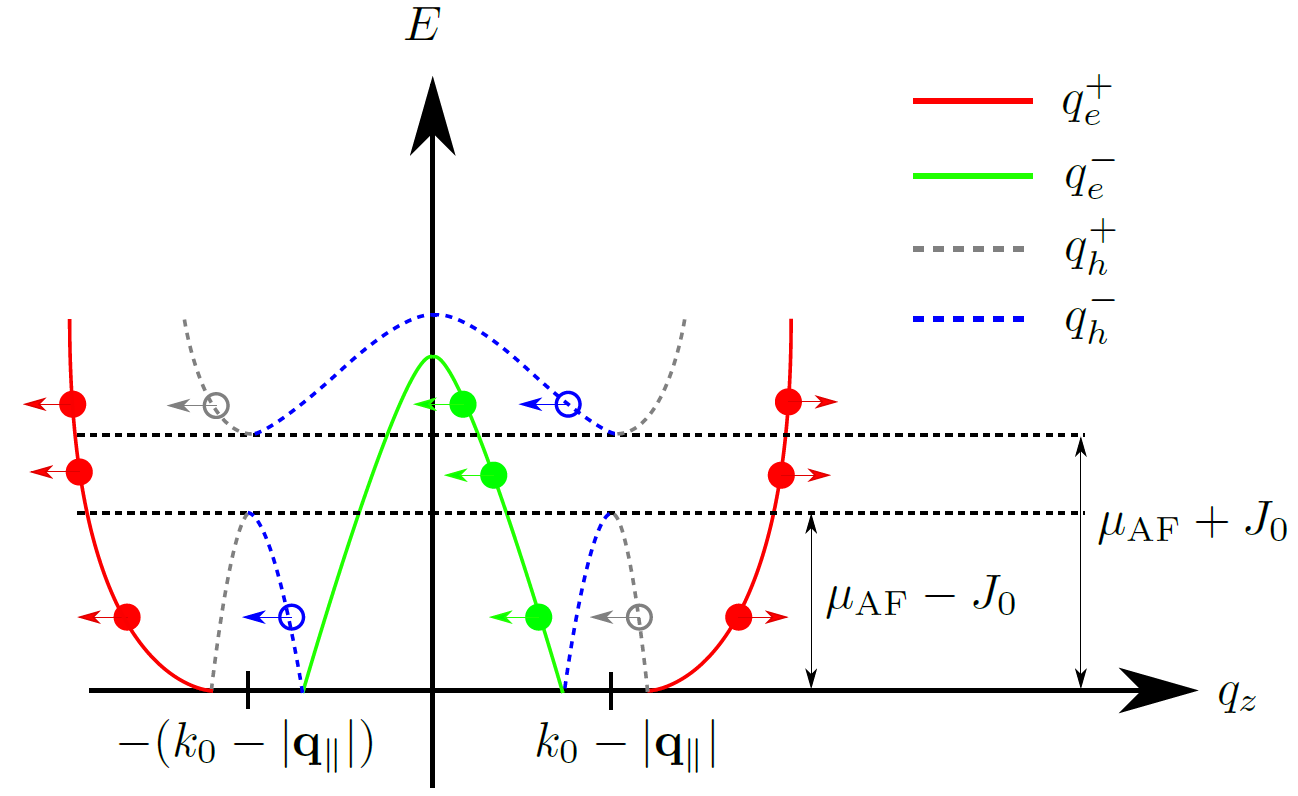}
\caption{$a)$ The allowed scattering processes, see Eq.~\eqref{Eq:Eigenvalues}. Electrons (holes) are drawn as filled (empty) circles. Incoming (reflected) particles are represented by rightward (leftward) arrows. $b)$ A sketch of the possible scattering processes that can occur at an AF--S junction.}
\label{fig:band_diagram}
\end{figure}

Figure \ref{fig:band_diagram}, shows a plot of the dispersion relations of the AF layer given in Eq.~\eqref{Eq:Eigenvalues}, where the possible scattering processes are identified. In contrast to an N(F)--S junction, an AF--S junction permits both specular AR and retro normal reflection (NR)~\cite{PhysRevLett.94.037005,PhysRevB.72.184510,PhysRevResearch.1.033095}.

In the following, we show how these two new scattering mechanisms, i.e., retro NR and specular AR, affect the transport properties of AF--S junctions.

\textit{Thermoelectric coefficients}.-- To study electrical and thermal transport, we assume that the AF is in contact with a biased reservoir, and that the S is in contact with a reference reservoir. Applying a bias voltage $U$ or a temperature difference $\Delta T$ through the junction induces an electric current or a heat current, respectively. In the BTK formalism~\cite{PhysRevB.25.4515,PhysRevB.48.15198,PhysRevB.51.16936}, the differential charge ($G_C=\dd I/\dd U$) and heat ($L_Q=\dd I/\dd \Delta T$) conductances read
\begin{subequations}
\begin{align}
        &G_C =\frac{ Ae^2 }{ 4\pi^3 \hbar} \int \dd E\, \dd^2\mathbf{q_{\parallel}} \frac{1-R_e + R_h}{4 k_{\mathrm{B}}T \cosh^2\left(\frac{E-eU}{2k_{\mathrm{B}}T}\right)},\label{Eq:conductances1}\\
        &L_Q = \frac{ A k_{\mathrm{B}} }{ 4\pi^3 \hbar} \int \dd E\, \dd^2\mathbf{q_{\parallel}} 
        \frac{E^2 \left(1-R_e-R_h\right)}{\left[2 k_{\mathrm{B}}\left(T+\Delta T\right)\right]^2 \cosh^2\left[\frac{E}{2k_{\mathrm{B}}\left(T+\Delta T\right)}\right]},
        \label{Eq:conductances2}
\end{align}
\end{subequations}
where $A$ and $T$ are the interfacial area and the thermal equilibrium temperature, respectively. The total reflection probabilities for electrons $(e)$ and holes $(h)$ are
\begin{align}
    &R_{e(h)}= \sum_{s} \left(R_{e(h),s}^+ + R_{e(h),s}^-\right).
\label{Eq:enhanceFactor}
\end{align}
Here, $R_{c,s}^{\pm}$ is the reflection probability for particles with wavevector $q_c^{\pm}$, where $c = e,h$ and $s = \uparrow, \downarrow$~\cite{supplement}. AR results in a net charge transfer of $2e$, but zero heat transfer~\cite{PhysRevLett.91.137002,doi:10.1098/rsta.2018.0140,PhysRevB.77.132503} across the interface; thus, AR increases the electrical conductance and decreases the thermal conductance.

\textit{Numerical parameters}.-- Before presenting our numerical results, we introduce our dimensionless parameters: the spin-independent barrier strength $Z = V_0 m/\hbar^2 q_{*}$, the Rashba strength $\lambda = 2\lambda_0  m/\hbar^2 q_{\mathrm{*}}$, and the exchange strength $J_0/\mu$. Here $q_*^2 = k_0^2 + q_{\mathrm{F}}^2$, where $q^2_{\mathrm{F}}=2m\mu/\hbar^2$. For simplicity, we set $\mu_{\mathrm{AF}} = \mu_{\mathrm{S}} = \mu$ and normalize the electrical and thermal conductance with respect to the corresponding Sharvin conductance~\cite{RevModPhys.76.323}: $\Tilde{G}_C = G_C/G_C^{\mathrm{Sh}}$ and $\Tilde{L}_Q = L_Q/L_Q^{\mathrm{Sh}}$. The Sharvin electrical (thermal) conductance is the electrical (thermal) conductance evaluated in the limit $\Delta_0 = J_0 = Z =0$, i.e., the response functions of a normal metal with perfect transmission: $G_C^{\mathrm{Sh}} = e^2 q_*^2 A / 4\pi^2 \hbar $ and $L_Q^{\mathrm{Sh}} = A k_{\mathrm{B}}^2 T_c q_*^2/12\hbar$.

In our calculations, we estimate the effective mass to be $\hbar^2 / 2m = 0.5\, \mathrm{eV}\, \mathrm{nm}^2$ based on a tight-binding model with typical material parameters~\cite{PhysRevLett.120.197202,PhysRevB.95.014403,PhysRevLett.113.157201}. Furthermore, the superconducting gap $\Delta_0$ is several orders of magnitude smaller than the chemical potential $\mu$. For concreteness, we set $\mu = 2\, \mathrm{eV}$ and allow the exchange strength to lie in the interval $ 0< J_0/\mu < 1$, where the system is conducting. As $J_0/\mu \rightarrow 1$, the AF material becomes an insulator, and the transport properties vanish. We consider the temperature range $0< T/T_c < 1$ so that superconductivity does not break down.

\begin{figure}[t]
\centering
\includegraphics[width=\columnwidth]{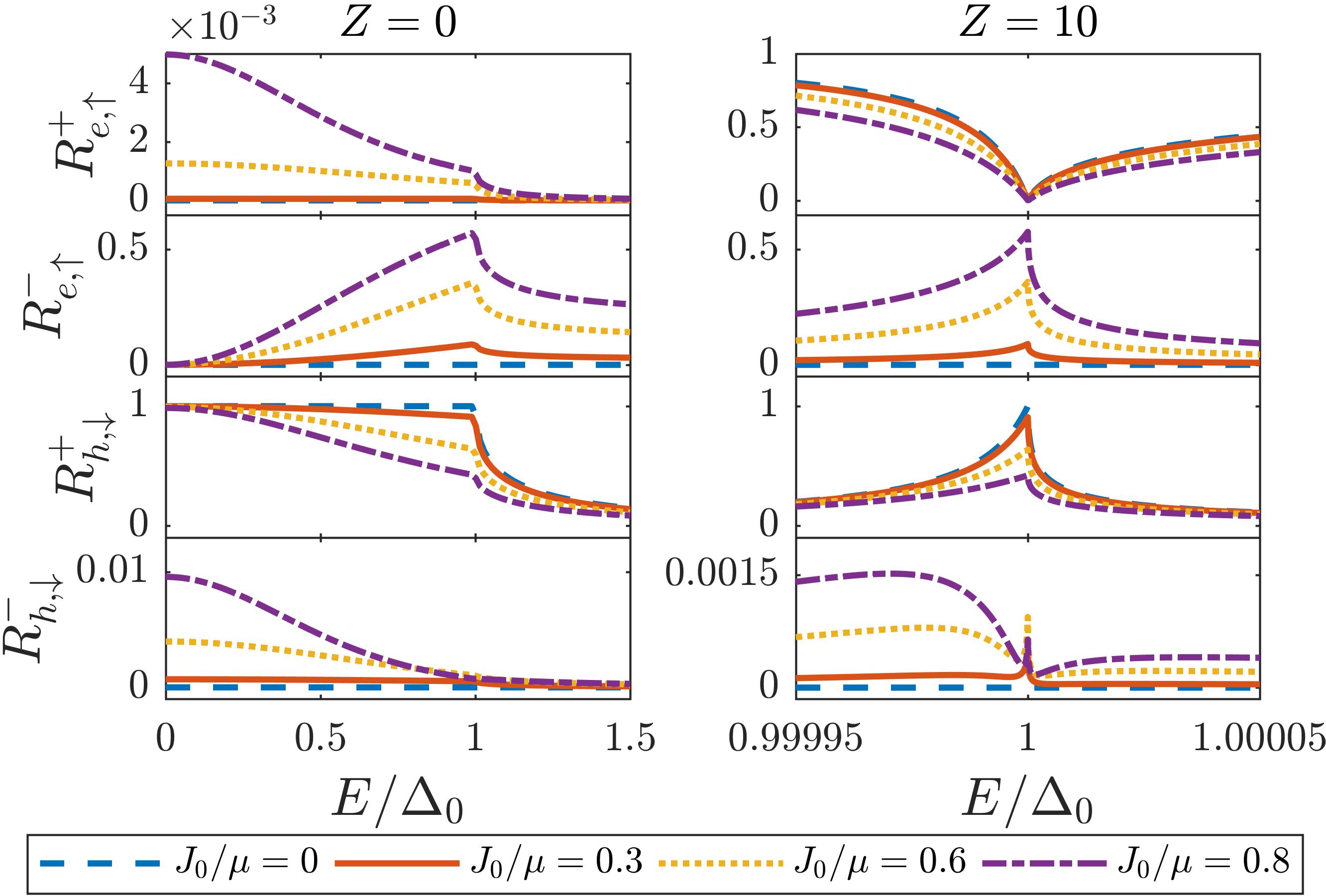}
\caption{The reflection probabilities as functions of the energy $E/\Delta$, the barrier strength $Z$, and the exchange strength $J_0/\mu$. The scattering processes associated with $R_{e,h}^-$ are absent at an N(F)--S junction; they are the result of the additional degrees of freedom in an AF. $R_e^-$ and $R_h^-$ correspond to retro NR and specular AR respectively.}
\label{fig:probs}
\end{figure}

\textit{Calculation of reflection probabilities}.-- Figure~\ref{fig:probs} shows the behavior of the reflection probabilities as functions of energy for different exchange strengths in both the transparent ($Z=0$) and tunneling ($Z \gg 1$) regimes in the absence of RSOC.

For simplicity, we first consider a transparent interface ($Z = 0$) and the subgap regime ($E<\Delta_0$). In the normal metal limit ($J_0 = 0$), we find that retro AR is the dominant scattering process~\cite{PhysRevB.25.4515}. Retro NR and specular AR increase as the exchange interaction $J_0$ increases. Because with the onset of $J_0$, the new scattering channels associated with the sublattice degrees of freedom become available. In the supergap regime ($E>\Delta_0$), electron-like and hole-like charge carriers can propagate in the S layer.

If the interface is not transparent ($Z \neq 0$), AR is suppressed while NR is enhanced, because fewer electrons are allowed to enter the S layer to form Cooper pairs. Increasing $J_0$ leads to an increase in retro NR and a decrease in specular NR, see Fig.~\ref{fig:probs}.

\begin{figure}[h]
\centering
\includegraphics[width=\columnwidth]{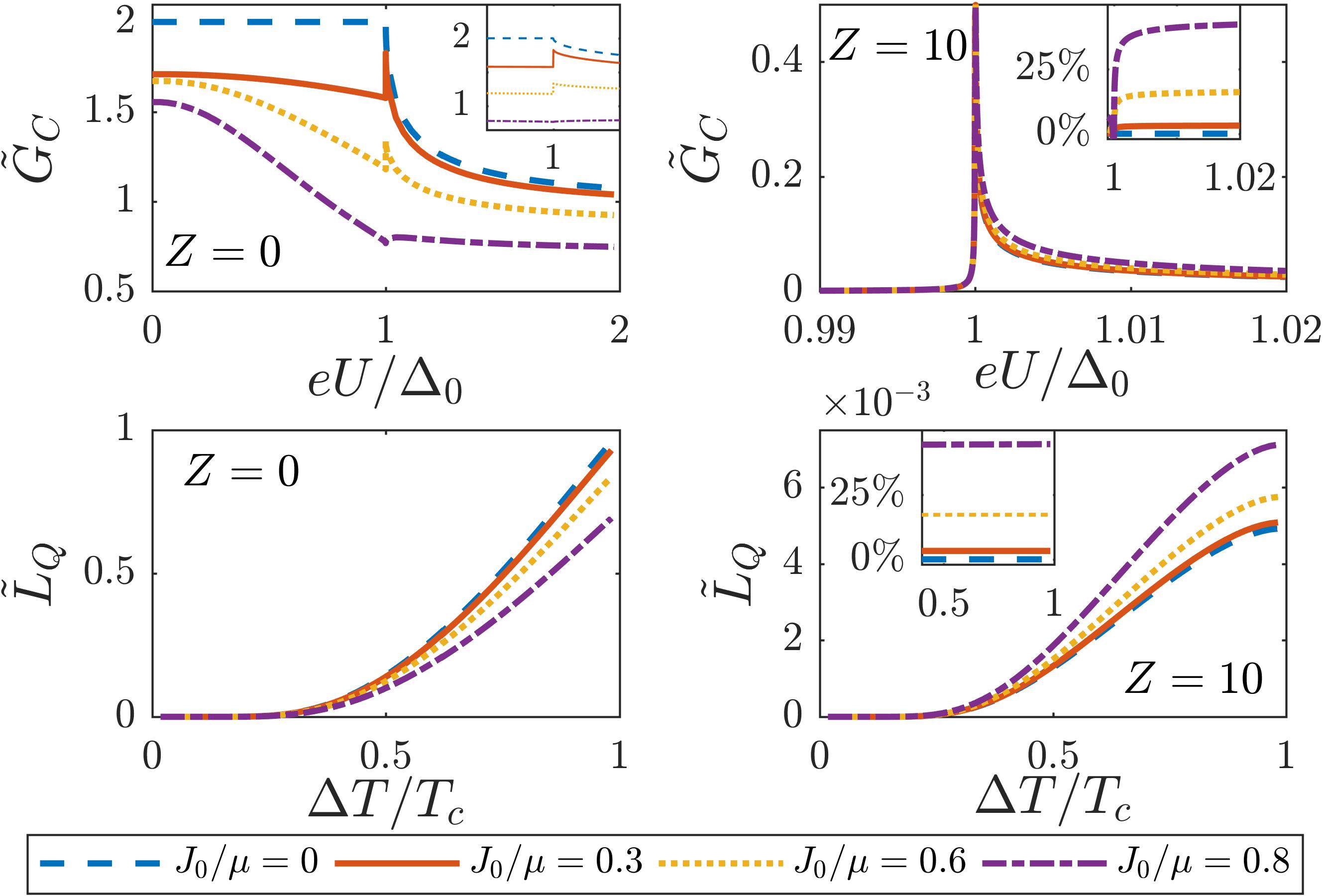}
\caption{The electrical conductance $\Tilde{G}_C$ and the thermal conductance $\tilde{L}_Q$ as functions of the dimensionless voltage $eU/\Delta_0$ and dimensionless temperature bias $\Delta T/T_c$, respectively, for different spin-independent barrier strengths $Z$ and exchange strengths $J_0/\mu$. The insets show the peak in the electrical conductance, and the percentage increase in the conductances as a function $J_0/\mu$ respectively.}
\label{fig:GcLqvsBias}
\end{figure}

\textit{Electrical and thermal conductance}.-- In the following, we elucidate experimental signatures in the response functions of the system. To simplify the discussion, we only consider the low temperature limit $T\rightarrow 0$ in Eqs. (\ref{Eq:conductances1}), and (\ref{Eq:conductances2}) in the rest of the paper. In Fig.~\ref{fig:GcLqvsBias}, we plot the electrical conductance and the thermal conductance as functions of the voltage and temperature bias, respectively, for different exchange and barrier strengths in the absence of SOC.

First, we focus on the electrical conductance shown in Fig.~\ref{fig:GcLqvsBias}. In the absence of a barrier and exchange interaction, the system behaves like a transparent N--S junction. In this case, each electron incident from the N layer enters the S layer and forms a Cooper pair, resulting in $100\%$ retro AR; consequently, the electrical conductance is $\tilde{G}_C = 2$. As the exchange strength increases, retro NR eventually becomes the dominant scattering process. Thus, with increasing $J_0$, less total charge is transported across the junction and the electrical conductance decreases. In contrast to a F--S junction, we find a sharp finite peak in the electrical conductance at $eU/\Delta_0 = 1$.

In the tunneling limit ($Z = 10$), the electrical conductance is singular at $eU/\Delta_0 = 1$, which originates from the singularity in the density of states (DOS) in the S layer.

In contrast to the electrical conductance, the thermal conductance is suppressed by AR. The physical reason is that Cooper pairs carry finite charge but zero heat across the junction. Therefore, for the thermal conductance to be finite, the temperature must be so high that electron-like and hole-like particles can be transmitted into the S layer. Since higher temperatures result in greater transmission of particles, the thermal conductance increases with increasing temperature bias, as shown in Fig.~\ref{fig:GcLqvsBias}.

In the transparent limit ($Z = 0$), the retro NR increases with increasing exchange strength. Since less particles are transmitted into the S layer, the thermal conductance decreases with increasing exchange strength. As the barrier strength increases, even fewer particles are transmitted into the S layer. In the tunneling limit ($Z = 10$), the thermal conductance is strongly suppressed.

Figure \ref{fig:GcLqvsBias} shows that, in the transparent limit, the increase of exchange strength reduces both the electrical and thermal conductance; by contrast, in the tunneling regime, the increase in exchange strength increases both of them. This behavior occurs due to the interplay between the exchange interaction and the barrier in the supergap regime ($E > \Delta_0$), where tunneling into the S layer is also allowed. In the tunneling limit, the exchange interaction enhances the transmission of both electron-like and hole-like particles into the S layer, consequently increasing the electrical and thermal conductance.

\begin{figure}[h]
\centering
\includegraphics[width=\columnwidth]{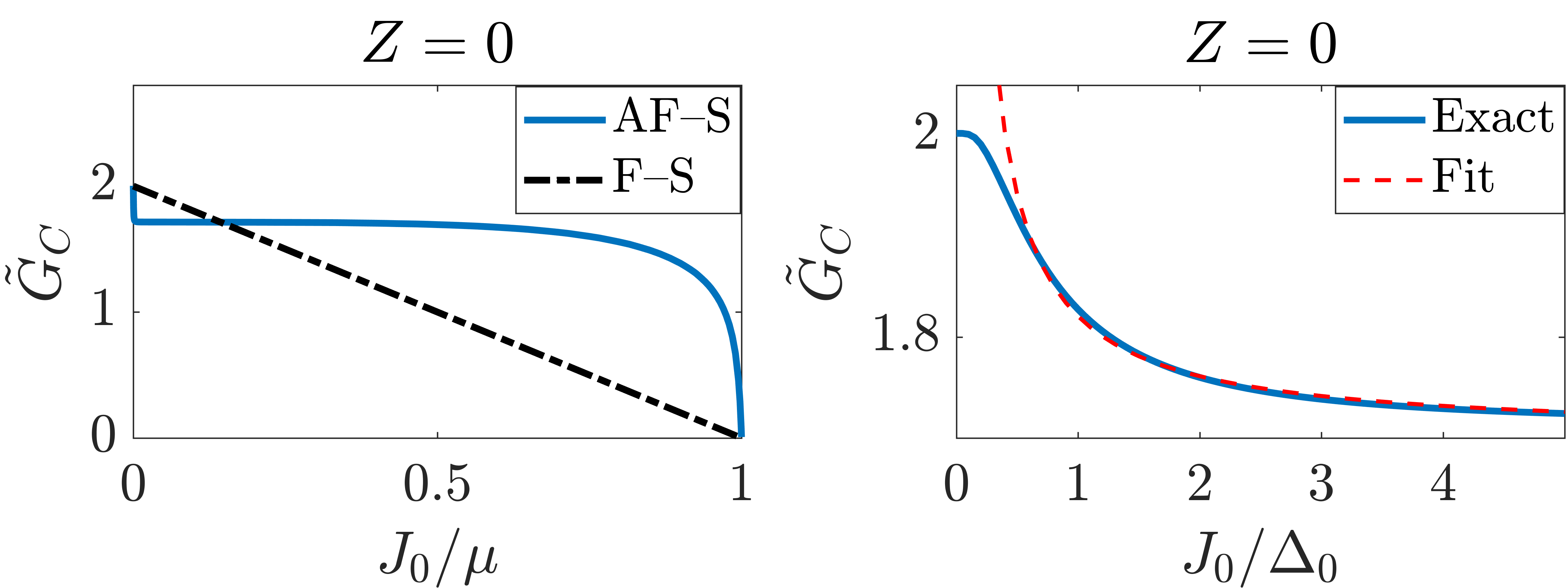}
\caption{\textit{Left:} zero-temperature electrical conductance $\Tilde{G}_C$ of the system as a function of the exchange strength $J_0/\mu$. The dash-dotted line represents the electrical conductance of an F--S junction~\cite{PhysRevLett.74.1657}. \textit{Right:} the behavior of $\Tilde{G}_C$ as a function $J_0/\Delta_0$. The dashed red line represents a numerical fit of the electrical conductance, $\tilde{G}_C\sim (J/\Delta_0)^{-1.0}$.}
\label{fig:GvsJ}
\end{figure}

To compare the AF--S junction with the F--S junction, we plot the electrical conductance as a function of the exchange strength in Fig.~\ref{fig:GvsJ}. In the F--S junction, the electrical conductance decreases linearly with increasing exchange strength, $\tilde{G}_C \approx 2(1-J_0/\mu)$~\cite{PhysRevLett.74.1657}. However, in the AF-S junction, the relationship between the electrical conductance and the exchange strength is more subtle. The electrical conductance decays rapidly at small $J_0/\mu$, is almost constant for intermediate $J_0/\mu$, and decays as $J_0/\mu \rightarrow 1$. We have checked that these features are robust by varying $m$, $\mu$, and $\Delta_0$ within the experimentally relevant intervals. The right panel of Fig.~\ref{fig:GvsJ} shows that the electrical conductance decays rapidly with increasing exchange strength on an energy scale set by the superconducting gap. In the regime where $J_0 \ll \Delta_0$, the system behaves like an N--S junction, such that $\tilde{G}_C = 2$. In the regime $J_0\sim\Delta_0$, we find that the reflection probabilities become dependent on the angle of incidence~\cite{supplement}. For electrons close to normal incidence, we find that retro AR dominates transport. For electrons with an angle of incidence nearly parallel to the interface, we find that retro AR is suppressed and specular NR is enhanced. This sudden enhancement of specular NR leads to the sharp decay of the electrical conductance observed in Fig.~\ref{fig:GvsJ}. Numerically, we find that $\tilde{G_c} \sim (J_0/\Delta_0)^{-1.0}$~\cite{supplement}.

In the regime $\Delta_0 \ll J_0 \ll \mu$, the DOS in the AF layer is approximately constant, and consequently, so is the electrical conductance~\cite{supplement}. As $J_0/\mu \rightarrow 1$, the AF layer starts to behave as an insulator, suppressing all transport properties.

\textit{Anisotropic magnetoresistance}.-- So far, we have not considered the effect of finite interfacial RSOC, resulting from the inversion symmetry breaking at the interface. For finite interfacial RSOC, additional scattering channels are opened in which spin-flip scattering is allowed. Spin-flipped AR allows for the formation of spin-triplet Cooper pairs in the S layer~\cite{PhysRevLett.115.116601,PhysRevB.101.014515,dutta2020thermoelectricity}. Recently, it has been found that in F--S junctions, interfacial RSOC leads to a large anisotropic magnetoresistance (AMR)~\cite{PhysRevLett.115.116601,PhysRevApplied.13.014030}, while there is no AMR in N--S junctions.

In the AF layer, the spin quantization axis is determined by the N\'eel vector. Consequently, a finite interfacial RSOC leads to anisotropy in the electrical and thermal conductance for an AF--S junction. Since we consider only an interfacial RSOC with an inversion-breaking axis in the $z$ direction, this AMR depends only on the N\'eel vector's polar angle $\theta$.

\begin{figure}[h]
\centering
\includegraphics[width=\columnwidth]{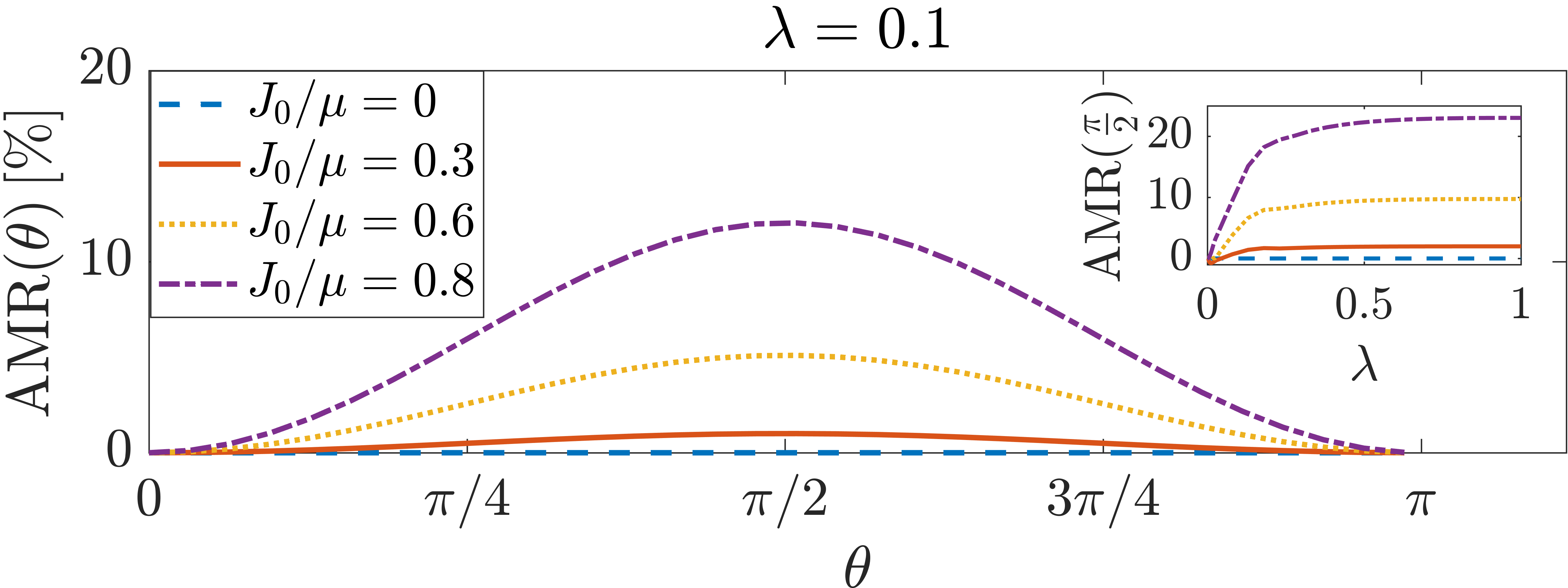}
\caption{The electrical AMR in an AF--S junction as a function of the orientation $\theta$ of the N\'eel vector and the exchange strength $J_0/\mu$. The inset show the dependence of the AMR maxima on the RSOC strength $\lambda$.}
\label{fig:soc}
\end{figure}

Figure~\ref{fig:soc} shows electrical $\mathrm{AMR}(\theta) = 1-\tilde{G}(0)/\tilde{G}(\theta)$ as a function of the N\'eel vector direction for a fixed RSOC strength. We find that the minima and maxima occur at $\theta = \{0, \pi\}$ and $\theta = \pi/2$, respectively. The inset shows that the maximum AMR increases with $\lambda$. The qualitative features of the electrical and thermal AMR is identical. Thus, similar to F--S junctions and in contrast to N--S junctions, AF--S junctions show a strong AMR. In an AF--N junction $(\Delta_0 \rightarrow 0)$, the electrical (thermal) AMR is approximately $75\%$ smaller ($50\%$ larger) than that in an AF--S junction. The simultaneous enhancement of the electrical AMR and diminution of the thermal AMR in an AF--S junction can be attributed to the finite AR in the presence of the S layer.

\textit{Concluding remarks}.--We demonstrate that the electrical and thermal conductance of AF--S junctions are qualitatively different from those of N(F)--S junctions due to the emergence of two new scattering processes: specular AR and retro NR. Furthermore, we show that there is a large AMR in the presence of a finite interfacial RSOC.

Our results reveal that superconducting spintronics based on antiferromagnetic materials, open a fascinating playground for novel physical phenomena. We hope that this theoretical study will inspire new experimental work on AF--S heterostructures.

\acknowledgments
This research was supported by the European
Research Council via Advanced Grant No. 669442,
``Insulatronics''; the Research Council of Norway through its Centres of Excellence funding scheme, Project No. $262633$, ``QuSpin'', and the Norwegian Financial Mechanism 2014-2021 Project No. 2019/34/H/ST3/00515, ``2Dtronics''.
P. D. was supported by the Science and Engineering Board (SERB) of the Department of Science and Technology (DST) (File No. PDF/2016/001178) of India and by the Swedish Research Council (Vetenskapsrådet Grant No. 2018-03488) and the Knut and Alice Wallenberg Foundation through the Wallenberg Academy Fellows Program. P. D. acknowledges the warm hospitality of the QuSpin group during her visit when this work was initiated.


\begin{thebibliography}{82}
\bibitem [{foo()}]{footnote}%
  \BibitemOpen
  \href@noop {} {\bibinfo  {journal} {In this paper, we have neglected the
  dependence of $T_c$ on the direction of the order parameter, typically
  originating from the strong SOC present in e.g., F--Heavy metal--S structures
  and the proximity effects~\cite{PhysRevB.102.020405}. We expect that $T_c$
  can depend on the orientation of $\mathbf{n}$ in AF--Heavy metal--S
  trilayers}\ }\BibitemShut {NoStop}%
\end{thebibliography}
%

\end{document}


\title{Supplemental Material:\\ Electrical and Thermal Transport in Antiferromagnet–Superconductor Junctions}

\author{Martin F.~Jakobsen}
\affiliation{Center for Quantum Spintronics, Department of Physics, Norwegian University of Science and Technology, NO-7491 Trondheim, Norway}
\author{Kristian B. Naess}
\affiliation{Center for Quantum Spintronics, Department of Physics, Norwegian University of Science and Technology, NO-7491 Trondheim, Norway}
\author{Paramita Dutta}
\affiliation{Department of Physics and Astronomy, Uppsala University, Box 516, S-751 20 Uppsala, Sweden}
\altaffiliation[Present address]{}
\affiliation{Institute of Physics, Sachivalaya Marg, Bhubaneswar 751005, India}
\author{Arne Brataas}
\affiliation{Center for Quantum Spintronics, Department of Physics, Norwegian University of Science and Technology, NO-7491 Trondheim, Norway}
\author{Alireza Qaiumzadeh}
\affiliation{Center for Quantum Spintronics, Department of Physics, Norwegian University of Science and Technology, NO-7491 Trondheim, Norway}

\maketitle

\section{Wavefunction in the antiferromagnet}
\begin{figure}[h!]
\centering
\includegraphics[width=0.7\columnwidth]{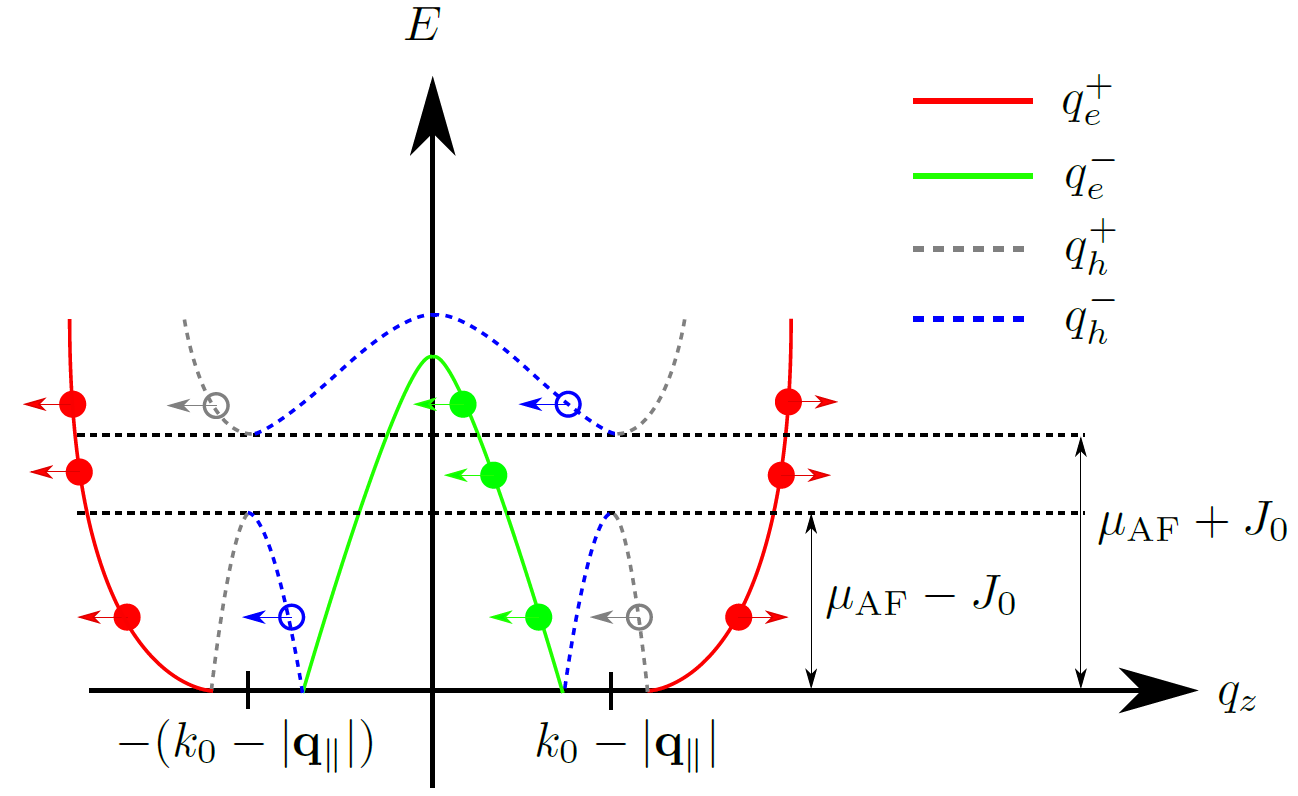}
\caption{The dispersion relation and the allowed scattering processes in the AF. Electrons (holes) are drawn as filled (empty) circles. The branches are parameterized as given in the figure. Incoming (reflected) particles are represented with a right (left) arrow. The new scattering processes not present in N(F)-S junctions are the ones labelled by $q_{e,h}^-$. The processes contributing to the conductances are the ones where $E \ll \mu_{\mathrm{AF}} - J_0$. Higher energies are suppressed by the Fermi-Dirac distribution. }
\label{fig:band_diagram}
\end{figure}

The AF--S junction is governed by the Bogoliubov de-Gennes (BdG) equation
\begin{equation}
    H \psi = E \psi,
    \label{Eq:ContinuumBdGEquation}
\end{equation}
as given in the main text.
Since the system is translationally invariant in the $xy$--plane the bulk eigenstates must be of the form 
\begin{equation}
    \psi = \chi e^{i\left(q_x x + q_y y\right)} e^{i q_z z},
    \label{Eq:ansatz}
\end{equation}
where $q_x$ and $q_y$ are unchanged upon reflection or transmission at the interface. By substitution of Eq. \eqref{Eq:ansatz} into Eq.~\eqref{Eq:ContinuumBdGEquation} for $z < 0$ we obtain
\begin{subequations}
\begin{eqnarray}
 q_z = q_{ e}^{\pm} = \sqrt{k_0^2 - \mathbf{q}_{\parallel}^2  \pm \frac{2m}{\hbar^2} \sqrt{\left(E + \mu_{\mathrm{AF}}\right)^2 - J^2}}, \\
 q_z = q_{ h}^{\pm} = \sqrt{k_0^2 - \mathbf{q}_{\parallel}^2  \pm \frac{2m}{\hbar^2} \sqrt{\left(E - \mu_{\mathrm{AF}}\right)^2 - J^2}},
\end{eqnarray}
\label{Eq:Eigenvalues}
\end{subequations}
where $\mathbf{q}_{\parallel} = (q_x,q_y)$, and 
\begin{equation}
    \begin{split}
        &\underline{q_z = q_e^+:}\\
        &\chi = \chi_{e,\uparrow}^+ = \left(E+\mu_{\mathrm{AF}}+b,a,\sqrt{\left(\mu_{\mathrm{AF}}+E\right)^2 - J_0^2 },0,0,0,0,0 \right),\\
        &\chi = \chi_{e,\downarrow}^+ =\left(a^*,\mu_{\mathrm{AF}}+E-b,0,\sqrt{\left(\mu_{\mathrm{AF}}+E\right)^2-J_0^2},0,0,0,0 \right),
    \end{split}
\end{equation}
\begin{equation}
    \begin{split}
        &\underline{q_z = q_e^-:}\\
        &\chi = \chi_{e,\uparrow}^- =\left(-(E + \mu_{\mathrm{AF}} + b),-a,\sqrt{\left(\mu_{\mathrm{AF}}+E\right)^2-J_0^2},0,0,0,0,0\right),\\
        &\chi = \chi_{e,\downarrow}^- =\left(-a^*,-\left(\mu_{\mathrm{AF}}+E-b\right),0,\sqrt{\left(\mu_{\mathrm{AF}}+E\right)^2-J_0^2},0, 0,0,0\right),
    \end{split}
\end{equation}
\begin{equation}
    \begin{split}
        &\underline{q_z = q_h^+:}\\
        &\chi = \chi_{h,\uparrow}^+ =\left(0,0,0,0,\mu_{\mathrm{AF}}-E+b,a^*,\sqrt{\left(\mu_{\mathrm{AF}}-E\right)^2-J_0^2},0\right),\\
        &\chi = \chi_{h,\downarrow}^+ =\left(0,0,0,0,a,\mu_{\mathrm{AF}}-E-b,0,\sqrt{\left(\mu_{\mathrm{AF}}-E\right)^2-J_0^2}\right),
    \end{split}
\end{equation}
\begin{equation}
    \begin{split}
        &\underline{q_z = q_h^-:}\\
        &\chi = \chi_{h,\uparrow}^- =\left(0,0,0,0,-\left(\mu_{\mathrm{AF}}-E+b\right),-a^*,\sqrt{\left(\mu_{\mathrm{AF}}-E\right)^2-J_0^2},0\right),\\
        &\chi = \chi_{h,\downarrow}^- =\left(0,0,0,0,-a,-\left(\mu_{\mathrm{AF}}-E-b\right),0,\sqrt{\left(\mu_{\mathrm{AF}}-E\right)^2-J_0^2}\right).
    \end{split}
\end{equation}
Here $a = e^{i \phi}J_0\sin\theta$ and $b = J_0 \cos\theta$. Unlike the eigenvectors of F in the F-S junctions the eigenvectors depend on the energy $E$, the chemical potential $\mu_{\mathrm{AF}}$, and the exchange interaction $J_0$~\cite{PhysRevLett.74.1657,PhysRevLett.115.116601,PhysRevB.96.115404}.
In Fig.~\ref{fig:band_diagram} we have plotted the branches $q_{e,h}^{\pm}$, and the scattering processes we are considering.

The wavefunction describing an incident electron from the AF is
\begin{equation}
        \psi_{\mathrm{AF}}=
        \begin{cases}
            e^{iq_e^+ z}\chi_{e,\uparrow}^+ + \sum_{\sigma} e^{-iq_e^+ z}r_{e,\sigma}^+ \chi_{e,\sigma}^+ + r_{e,\sigma}^- e^{iq_e^- z}\chi_{e,\sigma}^-+e^{ iq_h^+ z}r_{h,\sigma}^+ \chi_{h,\sigma}^+ + r_{h,\sigma}^- e^{- iq_h^- z}\chi_{h,\sigma}^-, & E < \mu_{\mathrm{AF}} - J_0  \\
            e^{iq_e^+ z}\chi_{e,\uparrow}^+ + \sum_{\sigma} e^{-iq_e^+ z}r_{e,\sigma}^+ \chi_{e,\sigma}^+ + r_{e,\sigma}^- e^{iq_e^- z}\chi_{e,\sigma}^-+e^{- iq_h^+ z}r_{h,\sigma}^+ \chi_{h,\sigma}^+ + r_{h,\sigma}^- e^{ iq_h^- z}\chi_{h,\sigma}^-, & E > \mu_{\mathrm{AF}} - J_0
        \end{cases}
        \label{Eq:wavefuncAF}
    \end{equation}
where $r_i^{\pm}$ represents the reflection amplitude for a particle being reflected to the branch $q_i^{\pm}$ with $i = e,h$. For small temperatures $T < T_c$ the Fermi-Dirac distribution makes sure that only the scattering processes with $E \ll \mu_{\mathrm{AF}} - J_0$ contributes to the electrical and thermal conductance. The sign in the exponents are determined by the $z$--component of the group velocity $v_z = \partial E / \partial q_z$ of the corresponding branch in Fig.~\ref{fig:band_diagram}.

\section{Calculating the reflection amplitudes}
By requiring that the wavefunction is continuous and integrating the BdG equation from $z = 0^-$ to $z = 0^+$ we obtain the matching conditions
\begin{equation}
\begin{split}
    &\psi_{\mathrm{AF}}\big|_{z = 0} = \psi_S\big|_{z = 0},\\
    &\frac{\hbar^2}{2m} \tau_z \otimes \sigma_x \otimes s_0 \left(\frac{d \psi_{\mathrm{AF}} }{dz}\Big|_{z= 0} - \frac{d \psi_{S} }{dz}\Big|_{z= 0}\right) =\left( V \tau_z \otimes \sigma_0 \otimes s_0 + \lambda_0 \tau_4^+ \otimes \sigma_x \otimes \left[\left(\bm{s} \times \bm{q}\right)\cdot \hat{z}\right]\right)\psi_S(0).
\end{split}
\end{equation}
Here $\psi_{\mathrm{AF}}$ and $\psi_S$ denotes the wavefunction in the AF and S respectively. We use these equations to calculate the reflection amplitudes in Eq. \eqref{Eq:wavefuncAF}.

\section{Probability and charge current}
Let $\Phi$ be an arbitrary eigenvector of the Hamiltonian. By defining the probability and charge density as $\rho_p = \Phi^{\dagger} \Phi$ and $\rho_C = e(|\Phi_1|^2+|\Phi_2|^2 + |\Phi_3|^2 + |\Phi_4|^2 - |\Phi_5|^2- |\Phi_6|^2 - |\Phi_7|^2 - |\Phi_8|^2)$ we evaluate $d\rho_{p,Q}/dt$ using the BdG equation to obtain the probability and charge currents
\begin{equation}
    \begin{split}
        &J_p = \frac{\hbar}{m} \sum_{i = 1, 2} \mathrm{Im}\left(\Phi_i^* \partial_z \Phi_{i+2} + \Phi_{i+2}^* \partial_z \Phi_{i} \right) -\frac{\hbar}{m} \sum_{j = 5, 6} \mathrm{Im}\left(\Phi_j^* \partial_z \Phi_{j+2} + \Phi_{j+2}^* \partial_z \Phi_{j} \right),\\
        &J_C = \frac{e\hbar}{m} \sum_{i = 1, 2} \mathrm{Im}\left(\Phi_i^* \partial_z \Phi_{i+2} + \Phi_{i+2}^* \partial_z \Phi_{i} \right) +\frac{e\hbar}{m} \sum_{j = 5, 6} \mathrm{Im}\left(\Phi_j^* \partial_z \Phi_{j+2} + \Phi_{j+2}^* \partial_z \Phi_{j} \right).
    \end{split}
\end{equation}
We use $J_p$ to evaluate the reflection probabilities $R_{i,\sigma}^{\pm}$ in the main text. The heat current is identical to the probability current except it includes a prefactor of $E-\mu$, for a detailed discussion see~\cite{PhysRevB.48.15198}. The charge and heat conductances in the main text are obtained from the charge and heat current by utilizing the BTK formalism~\cite{PhysRevB.25.4515,PhysRevB.48.15198}.
\section{The Normal Metal limit}
A crucial difference between the the N-S and AF-S junctions in our work is that the former has one physical lattice, while the latter has two physical sublattices. This results in that taking the limit $J_0 \xrightarrow{} 0$ can be subtle.

\begin{figure}[h!]
\centering
\includegraphics[width=\columnwidth]{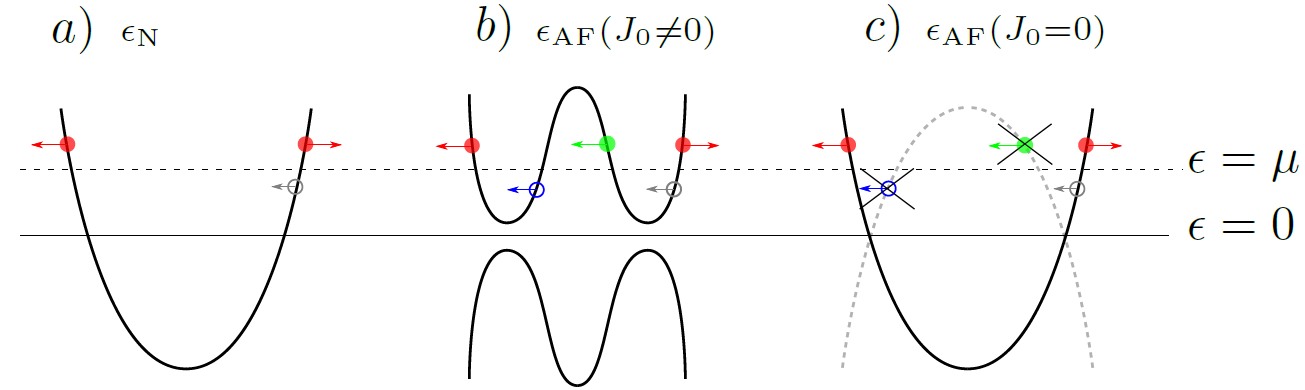}
\caption{The filled (empty) circles represents particles (holes). Incident (reflected) particles are represented by a right (left) arrow.
$a)$ The dispersion relation $\epsilon_{\mathrm{N}} = \frac{\hbar^2}{2m} (q^2- k_0^2)$ in N. Here only specular NR and retro AR is possible.
$b)$ The dispersion relation $\epsilon_{\mathrm{AF}} = \pm \sqrt{\left(\frac{\hbar}{2m}\left(q^2-k_0^2\right)\right)^2 + J_0^2}$ in AF, with two sublattices. In this case the additional scattering processes (green and blue) are allowed, because there are no redundancies in the degrees of freedom associated with the lattice. Physically the green and blue circles correspond to retro NR and specular AR.
$c)$ Naively taking the limit $J_0 \rightarrow 0$ gives the dispersion relations $\epsilon_{\mathrm{AF}} = \pm\frac{\hbar^2}{2m} (q^2- k_0^2)$ in the AF. Note that there is an additional band, due to the two sublattices compared to $a)$. The crossed out green (filled) and blue (empty) circles represents the scattering processes that are allowed in the AF-S junction but absent in the N-S junction.
} 
\label{fig:scatteringprocesses}
\end{figure}

In Fig.~\ref{fig:scatteringprocesses} $a)$ we have plotted the dispersion $\epsilon_{\mathrm{N}}  =\frac{\hbar^2}{2m} (q^2- k_0^2)$ which is valid in N. From the band diagram we can see that there are only two reflection processes that are possible: NR (red) and AR (gray).

In Fig.~\ref{fig:scatteringprocesses} $b)$ we have plotted the dispersions $\epsilon_{\mathrm{AF}} = \pm \sqrt{\left(\frac{\hbar}{2m}\left(q^2-k_0^2\right)\right)^2 + J_0^2}$ in the AF, where there are two sublattices. The gap is given by $2J_0$. From the band diagram we see that we obtain two additonal reflection processes indicated by the empty blue and filled green circles.

When we naively take the limit $J_0 \rightarrow 0$ we obtain the two bands $\epsilon_{\mathrm{AF}} = \pm\frac{\hbar^2}{2m} (q^2- k_0^2)$ shown in Fig.~\ref{fig:scatteringprocesses} $c)$. However, since the physics in Fig.~\ref{fig:scatteringprocesses} $a)$ and $c)$ should be the same it is clear that the band $\epsilon_{\mathrm{AF}} = -\frac{\hbar^2}{2m} (q^2- k_0^2)$ is unphysical. We confirm in the main text that the additional AR and NR associated with the unphysical band is zero in the limit $J_0 \rightarrow 0$.


\section{The Behavior of the Electrical conductance}
\begin{figure}[h!]
\centering
\includegraphics[width=0.7\columnwidth]{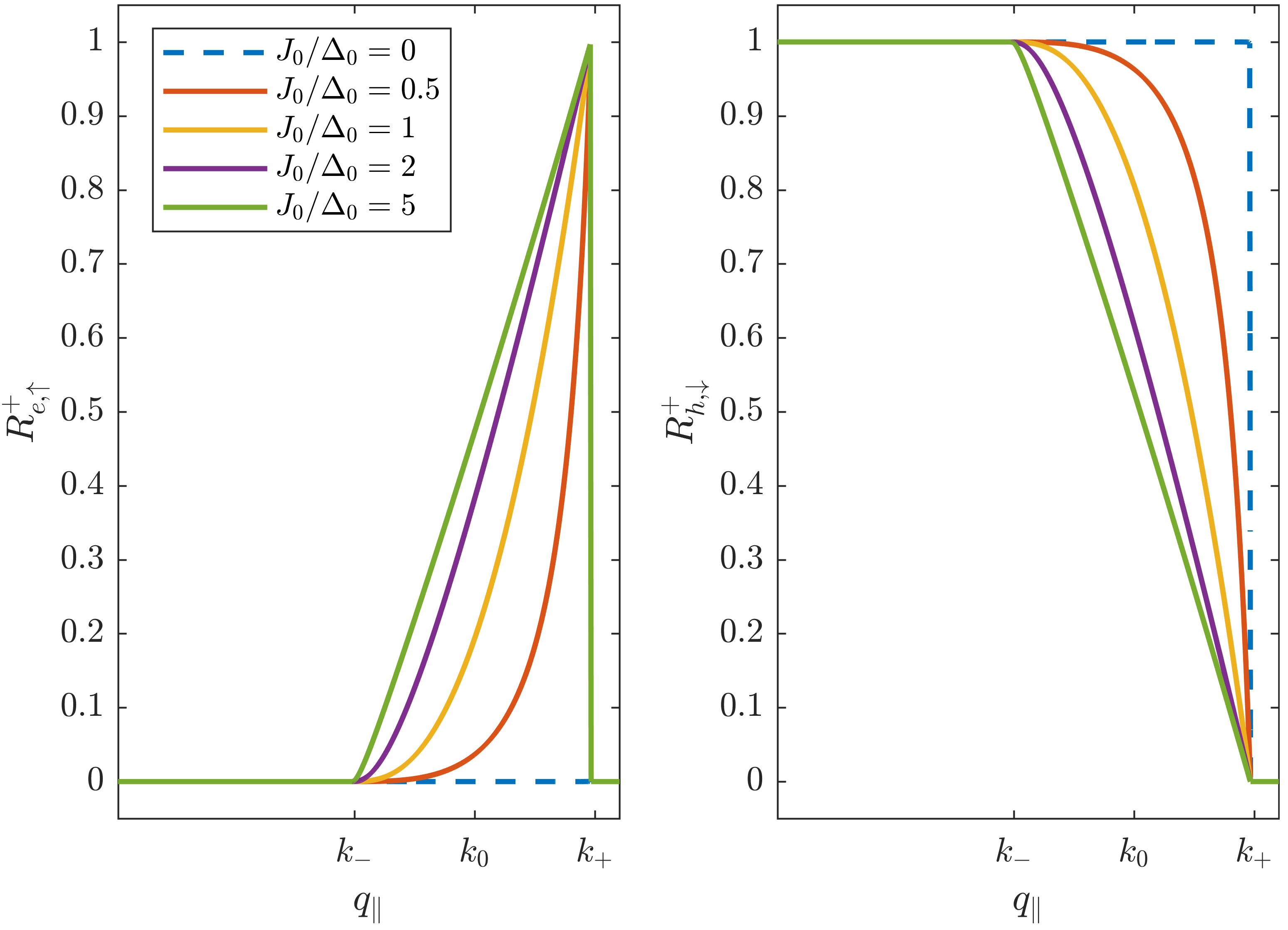}
\caption{Specular NR and retro AR as a function of $q_{\parallel}$. Note that in the regime $k_-<q_{\parallel}<k_+$, specular NR is enhanced and retro AR is suppressed when $J_0/\Delta_0 \neq 0$, even in the transparent limit $Z = 0$.}
\label{fig:critical_angles}
\end{figure}

In order to understand the dependence of the angle of incidence it is useful to employ the Andreev approximation, which utilizes that $E$ and $\Delta_0$ typically is much smaller than $\mu$ and $J_0$. In the AF the Andreev approximation reads $q_e^{\pm} \approx q_h^{\pm} \approx q_{\pm}$, where
\begin{equation}
\begin{split}
    &q_{\pm} = \sqrt{k_0^2 - \mathbf{q}_{\parallel}^2  \pm \frac{2m}{\hbar^2} \sqrt{\mu_{\mathrm{AF}}^2 - J_0^2}}.
\end{split}
\label{Eq:Eigenvalues_special}
\end{equation}
This equation result in two critical angles defined implicitly by the inequality
\begin{equation}
    \mathbf{q}_{\parallel}^2 < k_0^2 \pm \frac{2m}{\hbar^2} \sqrt{\mu^2 - J_0^2} \equiv k_{\pm}^2.
\end{equation}
We stress that the critical angle corresponding to $k_-$ plays no role in the N(F)-S junction because it originates from the unphysical dispersion $\epsilon_{\mathrm{AF}}\left(J_0 = 0\right) = - \frac{\hbar^2}{2m}(q^2-k_0^2)$, as discussed in the previous section. To elucidate the importance of the additional critical angle we have plotted the reflection coefficients as function of $q_{\parallel}$ and $J_0/\Delta_0$ in Fig.~\ref{fig:critical_angles} for $Z = 0$.

Note that as soon as $q_{\parallel} > k_-$ and $J_0 \neq 0$ the retro AR decays, while the specular NR increases. This behaviour results in the rapid decay of the electrical conductance shown in Fig. 4 in the main text. When $Z \neq 0$ this effect vanishes because the barrier enhances specular NR and suppresses retro AR. In the case $J_0 = 0$ the critical angle defined by $k_-$ plays no role in the scattering problem, and we obtain perfect retro AR.

Note that in the regime where $\Delta_0 \lessapprox J_0 \ll \mu$ we can approximate the retro AR as
\begin{equation}
    R_h^+=1-R_e^+\approx 
\begin{cases}
    1,& \text{if } q_{\parallel} < k_-\\
    \frac{k_+ - q_{\parallel}}{k_+ - k_-},& \text{if } q_{\parallel} \geq k_-.
\end{cases}
\end{equation}
This allows us to estimate the constant value of the zero-temperature conductance discussed in Fig. 4 in the main text. In this regime we obtain
\begin{equation}
    \Tilde{G}_C \approx \frac{2}{3}\left\{1 + \left(\frac{k_-}{k_+}\right) + \left(\frac{k_-}{k_+}\right)^2 \right\}\approx 1.71
\end{equation}
where we made the approximation $J_0 \ll \mu$ such that $k_{\pm} \approx k_0^2 \pm k_{\mathrm{F}}^2.$ We also provide an estimate for how quickly $\Tilde{G}_c$ decays as a function of $J/\Delta_0$, by performing a curve fit with the function
\begin{equation}
    \Tilde{G}_C \approx \alpha \left(J_0/\Delta_0\right)^\beta + \zeta.
    \label{Eq:CurveFit}
\end{equation}
The results are provided in Tab.~\ref{tab:CurveFit}.
\begin{table}[h]
	\centering
		\caption{The numerical values of the parameters in Eq. \eqref{Eq:CurveFit} for $\Tilde{G}_C$. The brackets $(\dots)$ give the $95\%$ confidence interval.} 
	\begin{tabular}{l l l} \hline \hline
	Parameter & Value & $95\%$-Interval  \\ \hline
    $\alpha$ & $0.118\, $ & $(0.117,0.119)$ \\ \hline
    $\beta$ & $-1.013\, $ & $(-1.025,-1.001)$ \\ \hline
    $\zeta$ & $1.703\, $ & $(1.702,1.703)$
    \\ 
    \hline \hline
	\end{tabular}
	\label{tab:CurveFit}
\end{table}


\section{Density of States}
\begin{figure}[h!]
\centering
\includegraphics[width=0.5\columnwidth]{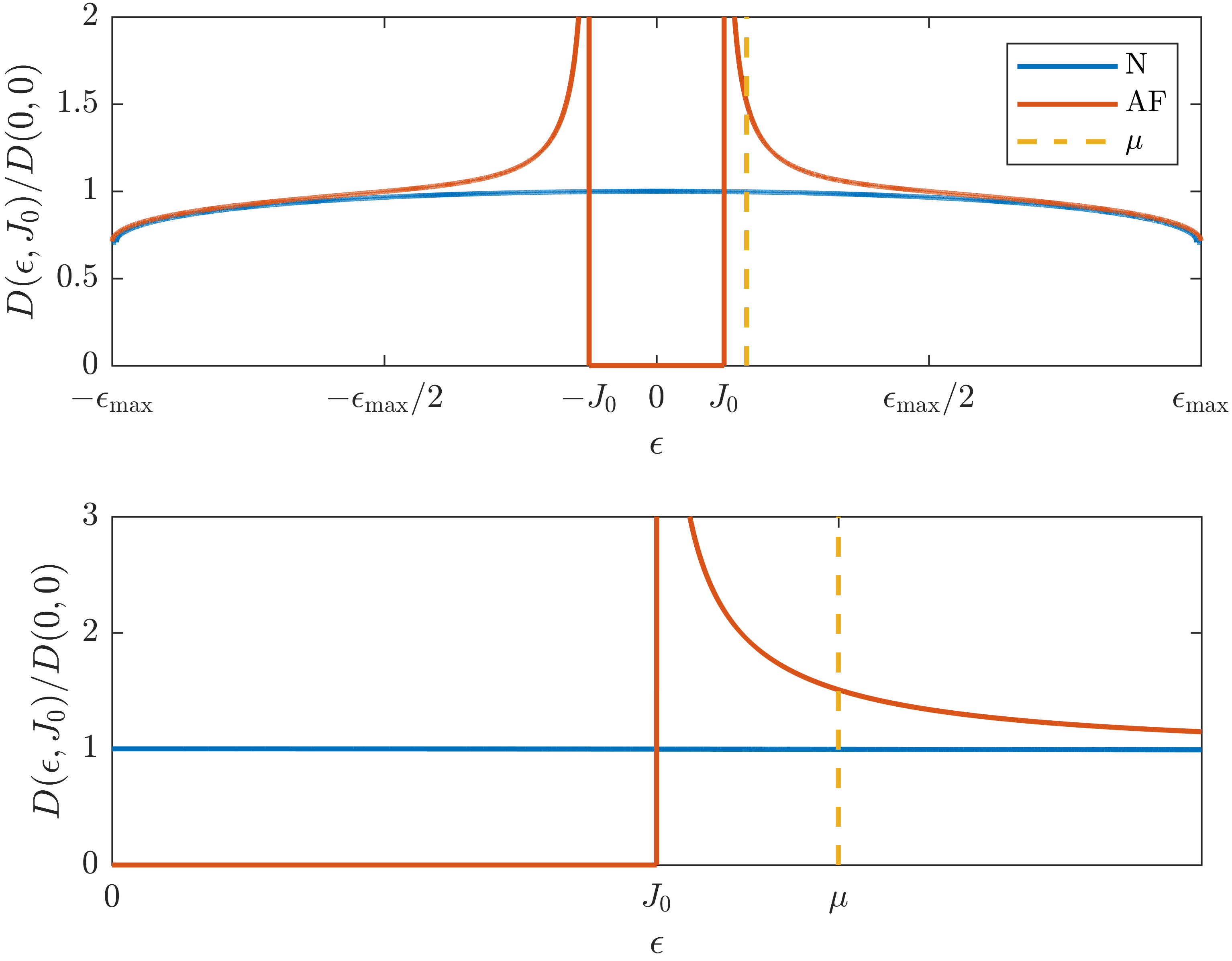}
\caption{DOS in N and AF. The maximum energy is $\epsilon_{\mathrm{max}} = \sqrt{\left(\hbar^2 k_0^2/2m\right)^2 + J_0^2}$. We have normalized the DOS with respect to its value in the normal metal at $\epsilon = 0$. When $J_0 \neq 0$ there is a strong asymmetry caused by the non-parabolic band in the AF.}
\label{fig:DosPlot}
\end{figure}

The density of states (DOS) in the AF is defined as
\begin{equation}
D(\epsilon) = \int \frac{d^3q}{(2\pi)^3} \delta(\epsilon - \epsilon_{\mathrm{AF}}(q)).
\end{equation}
In the AF $\epsilon_{\mathrm{AF}}(q) = \pm \sqrt{\left(\frac{\hbar^2}{2m}\left(q^2-k_0^2\right)\right)^2 + J_0^2}$ which gives
\begin{equation}
    \begin{split}
        D(\epsilon,J_0) = \frac{\Theta(|\epsilon| - J_0)}{2 \pi^2} \frac{m}{\hbar^2} \frac{|\epsilon|}{\sqrt{\epsilon^2 - J_0^2}}
        \left[ \sqrt{k_0^2 + \frac{2m}{\hbar^2} \sqrt{\epsilon^2 - J_0^2}} + \sqrt{k_0^2 - \frac{2m}{\hbar^2} \sqrt{\epsilon^2 - J_0^2}}\right].
    \end{split}
\end{equation}
In Fig.~\ref{fig:DosPlot} we have plotted the DOS in N and AF for $J_0/\mu = 0.75$. Note that the divergence at $\epsilon = J_0 \neq 0$ induces a strong asymmetry around the chemical potential, which increases with $J_0$, in the DOS. The asymmetry comes from the additional curvature of the dispersion in the AF, that we have sketched in Fig.~\ref{fig:scatteringprocesses} b). This growing asymmetry results in that AR is reduced when $J_0$ increases. We emphasize that the asymmetry present in the AF is fundamentally different from the one present in F, which is induced by the exchange spin-splitting of the energy bands. For energies $\epsilon < J_0$ the system behaves as a AF insulator, and no carriers are allowed to move.
\pagebreak
\section{Individual AMR contributions}
\begin{figure}[h]
\centering
\includegraphics[width=\columnwidth]{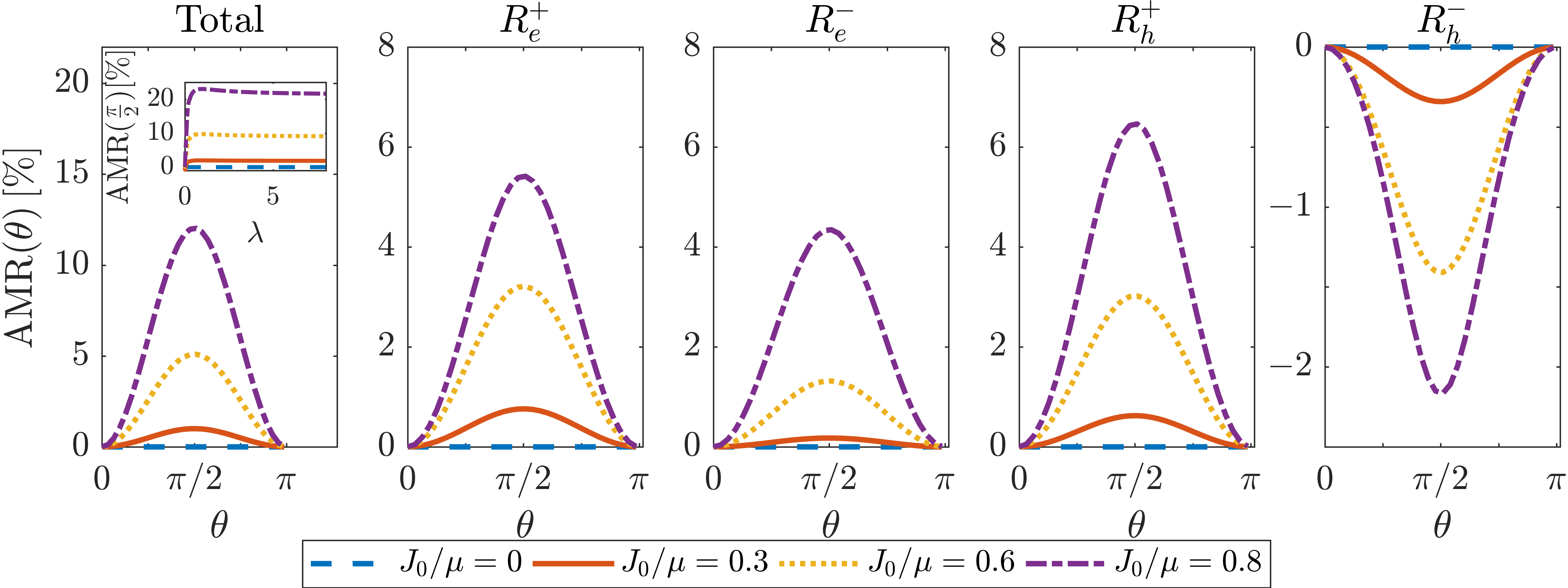}
\caption{The total and individual contributions to electrical AMR in an AF--S junction as a function of the orientation $\theta$ of the N\'eel vector and the exchange strength $J_0/\mu$. The inset show the dependence of the AMR maxima on the RSOC strength $\lambda$.}
\label{fig:soc}
\end{figure}

In Fig.~\ref{fig:soc} we have resolved the total AMR, given in the main text, into its individual contributions.
The specular NR and retro AR increases the AMR. The additional scattering processes retro NR and specular AR increases and decreases the AMR respectively. The AMR occurs due to the complicated interplay between the interfacial Rashba spin-orbit coupling and the direction of the Néel vector in the AF.

We find that the the new scattering processes $R_{e,h}^-$ suppresses the conductance. This is another essential difference when compared to the AMR in F--S junctions, and might lead to experimentally well-distinct variations of the AMR amplitudes.


%
